\documentstyle[12pt,aasms4]{article}   

\def\msun{{\rm M_{\odot}}}
\def\rsun{{\rm R_{\odot}}}

\def\be{\begin{equation}}
\def\ee{\end{equation}}
\begin{document}

\title{THE EVOLUTIONARY STATUS OF SS433}

\author{
Andrew~R.~King\altaffilmark{1, 2},
Ronald~E~Taam\altaffilmark{1, 3},
Mitchell~C.~Begelman\altaffilmark{1, 4}}

\authoremail{ark@star.le.ac.uk, taam@apollo.astro.nwu.edu,  
mitch@jila.colorado.edu
}

\altaffiltext{1} {Institute for Theoretical Physics, University of
California at Santa Barbara, CA 93106-4030, U.S.A.}

\altaffiltext{2} {Astronomy Group, University of Leicester,
Leicester LE1 7RH, U.K.}

\altaffiltext{3}{Department of Physics and Astronomy, Northwestern
University, Evanston, IL 60208}

\altaffiltext{4}{JILA, University of Colorado, Boulder, CO 80309-0440,
U.S.A.}

\begin{abstract}
We consider possible evolutionary models for SS~433. We assume that
common--envelope evolution is avoided if radiation pressure is able to
expel most of a super--Eddington accretion flow from a region smaller
than the accretor's Roche lobe. This condition is satisfied, at least
initially, for largely radiative donors with masses in the range
$4\msun - 12\msun$. For donors more massive than about $5\msun$,
moderate mass ratios $q = M_2/M_1 \ga 1$ are indicated, thus
tending to favor black--hole accretors. For lower mass donors,
evolutionary considerations do not distinguish between a neutron star
or black hole accretor.  In all cases the mass transfer (and mass loss) rates 
$\dot M_{\rm tr} \sim 7\times 10^{-6} - 4\times 10^{-4}\ \msun$~yr$^{-1}$
are much larger than the likely mass--loss
rate $\dot M_{\rm jet} \sim 10^{-6}\ \msun$~yr$^{-1}$ in the
precessing jets. Almost all of the transferred mass is expelled at
radii considerably larger than the jet acceleration region,
producing the `stationary' H$\alpha$ line, the infrared luminosity,
and accounting for the low X--ray luminosity.

\end{abstract}


\section{INTRODUCTION}

The nature of the unusual binary system SS~433 has been an interesting
question ever since the recognition (Abell \& Margon, 1979; Milgrom,
1979; Fabian \& Rees, 1979) that the system drives
precessing jets with velocities $\simeq 0.26c$. In particular,
although the binary period $P = 13.1$~d has long been known (Crampton
\& Hutchings, 1981), there is no consensus about the component masses
since all radial--velocity studies have to use emission lines, and one
cannot be sure that the measured velocities are those of either
star. The heavy extinction towards the object makes estimates of the
companion's spectral type and luminosity difficult; dereddening with
the usual values $A_V \simeq 7$ based on comparison of infrared and
H$\alpha$ intensities (e.g. Giles et al., 1979) suggests the presence
of a Rayleigh--Jeans continuum in the optical, and thus possibly an
early--type companion. Estimating the mass ratio from the duration of
the observed X--ray eclipse (Kawai et al., 1989) is also difficult, as
we know that the X--rays come from the moving jets (Watson et al.,
1986) and may therefore be extended. The assumptions that the jets are
partly obscured by the accretion disc and by the companion during
eclipse lead (D'Odorico et al., 1991)
to estimates $q = M_2/M_1 \sim 4$, where $M_2, M_1$ are
the masses of the companion and compact star respectively,
consistent with the presence of an early--type companion.

Observational hints that SS~433's companion is relatively massive
($q \ga 1$) have, until recently, presented a dilemma to theorists. A
large mass ratio would put SS~433 in a state often invoked in binary
evolution scenarios. This situation tends to lead to high mass
transfer rates, with a large fraction of the donor's mass being
transferred on its thermal timescale if its
envelope is predominantly radiative, and even more rapidly if the
envelope is convective. The resulting short mass transfer lifetime
for a fairly massive donor (i.e., $q \ga 1$) offers a simple
explanation for the uniqueness of SS~433. Furthermore, a high mass
transfer rate is indicated by the mass loss rate
$\dot M_{\rm jet} \sim 10^{-6}\ \msun$~yr$^{-1}$
in the precessing jets (Begelman et al., 1980). However, this is also a
potential problem: if the companion is indeed relatively massive, even
thermal--timescale mass transfer leads to rates far in excess of the
Eddington limit for an accretor of a few $\msun$, in excess even of the
inferred mass-loss rates in the jets.
Conventional wisdom has until recently suggested that
such rates inevitably lead very quickly to common--envelope (CE)
evolution, in which the compact object can neither accrete nor
expel the transferred matter rapidly enough to prevent the formation
of an envelope around the entire system. This would then
probably appear as a giant, and certainly not be recognizable as an
accreting binary. Since SS~433 is not yet in such a state, the predicted
lifetime for its current state would become embarassingly short,
requiring very high space densities of similar systems.

The problem is only slightly
eased by abandoning the assumption $q \ga 1$, since the mean density
of the companion is essentially fixed by the requirement that it
should fill its Roche lobe in a binary with a period of
13.1~d (mass transfer through stellar wind capture is very unlikely to
give the high mass transfer rates inferred).
A companion star in the process of crossing the Hertzsprung
gap is the only likely possibility, as a main--sequence companion
would have to be improbably massive (see eqs \ref{4}, \ref{5} below)
while nuclear--timescale mass transfer from a giant companion would
give far too low a transfer rate. This then leads back to mass
transfer on something like the thermal time of the expanding companion
star; while this is somewhat milder than in the case $q \ga 1$, it
would still be well above the values hitherto thought likely to
produce a common envelope.

Recent work on the neutron--star X--ray binary Cygnus X--2 (King \&
Ritter, 1999; Podsiadlowski \& Rappaport, 1999) offers a way
out of this dilemma: it is evident that this system has survived an
episode of thermal--timescale mass transfer resulting from an initial
mass ratio $q_i \ga 1$ without entering CE
evolution. The aim of this paper is to investigate whether this is
possible in SS~433 also, and thus to discover its likely evolutionary
status.

\section{AVOIDANCE OF COMMON ENVELOPE EVOLUTION}

King \& Ritter (1999) show that the progenitor of Cygnus X--2 must
have transferred a mass $\sim 3\msun$ at rates
$\dot M_{\rm tr} \ga 10^{-6}\ \msun$~yr$^{-1}$ (the Case AB
evolution suggested by  Podsiadlowski \& Rappaport, 1999 leads to a
similar requirement). This
greatly exceeds the Eddington rate $\dot M_{\rm Edd} = L_{\rm
Edd}/c^2$ for a
$1.4\msun$ neutron star. This star evidently accreted only a tiny
fraction of the transferred mass, expelling the rest from the system
entirely. The obvious agent for this is radiation pressure.
King \& Begelman (1999) (hereafter KB99)
suggest that expulsion occurs from the `trapping' radius
\be
R_{\rm ex} \sim \biggl({\dot M_{\rm tr}\over \dot M_{\rm Edd}}\biggr)R_S
\simeq 1.3\times 10^{14}\dot m_{\rm tr}~{\rm cm},
\label{1}
\ee
where $R_S$ is the Schwarzschild radius and $\dot m_{\rm tr}$ is the
transfer rate expressed in  $\msun$~yr$^{-1}$. (Note that $R_{\rm ex}$
is independent of $M_1$ since both $\dot M_{\rm Edd}$ and $R_S$ scale
as $M_1$.) Within this radius advection drags photons inward, 
overcoming their outward diffusion through the
matter. If the matter has even a small amount of angular momentum most
of it is likely to be blown away as a strong wind from $R_{\rm ex}$:
the gravitational energy released by accretion at $\sim \dot M_{\rm Edd}$
deep in the potential of the compact star is used to expel the remainder
($\sim \dot M_{\rm tr}
{\gg}
\dot M_{\rm Edd}$) of the infall,
which is only weakly
bound at distances $\sim R_{\rm ex}$ (cf Blandford \& Begelman
1999). KB99 suggest that CE evolution is avoided provided that
$R_{\rm ex}$ is smaller than the accretor's Roche lobe radius $R_1$.
If the accretor is the less massive star this is given approximately by
\be
R_1 = 1.3\times 10^{11}m_1^{1/3}P_{\rm d}^{2/3}\ {\rm cm},
\label{2}
\ee
where $m_1 = M_1/\msun$ and $P_{\rm d}$ is the binary period in days.
Since this is related to the Roche lobe radius $R_2$ of the companion
star via
\be
{R_2\over R_1} \simeq \biggl({M_2\over M_1}\biggr)^{0.45},
\label{3}
\ee
(cf King et al., 1997) KB99 were able to determine whether Roche--lobe
overflow from various types of companion star
would lead to CE evolution or not. They concluded that
CE evolution was unlikely for
mass transfer from any main--sequence or
Hertzsprung gap star with $q\ga 1$ provided that its envelope was
largely radiative. In the next section
we investigate possible companion stars in SS~433.

\section{THE EVOLUTION OF SS~433}

If the companion star in SS~433 is more massive than the
compact accretor ($q > 1$) and fills its Roche lobe, then from
(\ref{2}) and (\ref{3}) it has radius
\be
R_2 = 10\rsun m_1^{-0.12}m_2^{0.45}\biggl({P_{\rm d}\over
13.1}\biggr)^{2/3}.
\label{4}
\ee
In the opposite case ($q\la 1$) we have instead
\be
R_2 = 10\rsun m_2^{0.33}\biggl({P_{\rm d}\over 13.1}\biggr)^{2/3}.
\label{5}
\ee
In either case this is obviously a fairly extended star, and clearly
well above the upper main sequence mass--radius relation for any
realistic mass $M_2 = m_2\msun$. By the
reasoning of the previous section, the only possible companions are
stars which came into contact with the Roche lobe as they crossed the
Hertzsprung gap. Since such stars are by definition out of thermal
equilibrium, we cannot assume that their structure is given by that of
a single star of the same instantaneous mass (although this is
approximately true for companions of modest mass $m_2 \la 3.5$ with $q
\la 1$; Kolb, 1998). Instead we must follow their evolution under mass
loss explicitly.

The lack of clear dynamical mass information means that there is
considerable freedom in trying to fit the current state of SS~433, and
one cannot expect to find a unique assignment. We restricted the
evolutions we considered to those satisfying the following list of
conditions at the current epoch:
\bigskip

1. $P_d \simeq 13.1$
\bigskip

2. Mass transfer rate $\dot M_{\rm tr} >
\dot M_{\rm jet} \sim 10^{-6}\ \msun$~yr$^{-1}$
\bigskip

3. $R_{\rm ex} < R_1$
\bigskip

4. The companion does not have a deep convective envelope (which would
make CE evolution inevitable). In practice this means that the stellar
effective temperature should typically exceed a value $\sim 6000$~K
which in turn requires an initial companion mass $M_{2i}\ga 4\msun$
\bigskip

5. The time since mass transfer exceeded the Eddington limit is 
$t_0 \sim 10^3$~yr.
\bigskip

Condition 5 comes from observations of the surrounding W50 nebula.
If this is attributed to interaction with the jets one finds $t_0 \sim
10^3$~yr, assuming that the jets were produced promptly
(Begelman et al., 1980; K\"onigl, 1983).
We calculated evolutionary models satisfying these conditions
using the code developed by Eggleton (1971, 1972), with the Roche lobe
radius given (more accurately than \ref{4}, \ref{5} above) by
\be
{R_2\over a} = {0.49q^{2/3}\over 0.6q^{2/3} + \ln(1+q^{1/3})},
\label{9}
\ee
with $a$ the binary separation and a mass transfer formulation as 
described in Ritter (1983). In all cases the mass transfer rate 
is highly super--Eddington; we assume that the transferred mass is
lost from the binary with the specific angular momentum of the compact
accretor. From eq. 5 of King \& Ritter (1999) it is easy to show that
the orbital period decreases for mass ratios $q > q_{\rm crit} \simeq
1.39$, and increases for smaller $q$.
Because higher mass stars drive higher rates of mass transfer,
the companion star mass is limited
to $\la 12 \msun$ for otherwise birth rate requirements for systems
similar to SS~433 would become severe (see below). Table 1 shows our
results
for systems characterized by an initial orbital period of 13.1 d.

\section{DISCUSSION}

Table 1 provides only a coarse sampling of the parameter space of
possible evolutionary models for SS~433. However we can already make
some interesting statements.

For a given initial mass {\it ratio} $q_i = M_{2i}/M_1$, (e.g. sequences
2
and 4) higher masses imply higher mass loss rates, and hence that more
mass is lost after a given time ($\sim 10^3$~yr). We can understand
this as a consequence of the shorter thermal timescale for
higher--mass stars. The orbital period
decrease is greater for higher masses as a result. (Note that sequence
1 has $q_i$ very close to $q_{\rm crit}$.) For a given initial
companion mass $M_{2i}$ (compare sequences 1 - 3) decreasing the
compact object mass produces the same trends, because it amounts to
increasing the mass ratio further above $q_{\rm crit}$.
For a neutron--star accretor (sequence 5) the larger
mass ratio wins out over the smaller companion mass, again producing
the same trends. This implies that a neutron--star accretor provides a
good fit to the present state of SS~433 only if the companion has
quite a low mass ($\sim 4 - 5\msun$) in a fairly narrow range: masses
much lower than this imply large convective mass fractions at or soon
after the start of mass transfer, and thus CE evolution. This
results in a dynamical instability or a delayed dynamical
instability as discussed by Webbink (1977), Hjellming \& Webbink (1987),
and Hjellming (1989).
For companion stars more massive than about $5 \msun$ with a neutron
star companion, binary evolution leads to a
decrease of the orbital separation and period
even after 1000 yrs, excluding such systems as candidate progenitor
systems for SS~433. For longer initial periods, the companion is
likely to have a convective envelope at the onset of mass transfer,
and thus enter a CE stage. 

Depending on the component masses, mass transfer rates 
$\dot M_{\rm tr} \sim 7\times 10^{-6} - 4\times 10^{-4}\ \msun$~yr$^{-1}$
are typical for a system
with a period of 13.1~d, and can evidently be ejected without causing
the onset of CE evolution provided that the companion is predominantly
radiative.  The resulting mass transfer lifetimes are in the range
$10^4 - 10^5$~yr, given largely by the thermal timescale of the
companion.  They cannot be much greater than this, since this requires
companion masses $\la 4\msun$, which are subject to
a dynamical instability. The birthrate requirement for systems like
SS~433 is thus of order $10^{-4} - 10^{-5}$~yr$^{-1}$ in the Galaxy.

The predicted mass transfer rates 
$\dot M_{\rm tr} \sim 7\times
10^{-6} - 4\times 10^{-4}\ \msun$~yr$^{-1}$ 
are in all cases much larger than the likely mass--loss
rate $\dot M_{\rm jet} \sim 10^{-6}\ \msun$~yr$^{-1}$ in the
precessing jets (Begelman et al., 1980). We expect that the jets are ejected 
from a region no larger than $R_{\rm jet} \sim (\dot M_{\rm jet}/\dot
M_{\rm tr})R_{\rm ex} \sim 1\times 10^8$~cm; 
their quasi-relativistic velocity suggests that they emerge from a
smaller radius, i.e., a few times the
neutron--star radius $10^6$~cm or the black--hole Schwarzschild radius
$3\times 10^5m_1$~cm.
The jets constitute just the innermost part of
the mass expulsion from the accretion flow: almost all of
the transferred mass is lost from larger radii, $\ga R_{\rm ex}$.  In
addition to accounting for the very low observed X--ray luminosity
(which probably comes entirely from the jets), this expulsion of matter
is presumably the source of the `stationary' H$\alpha$ line and the
associated free--free continuum seen in the near infrared (Giles et
al., 1979). The emission measure $VN_e^2 \simeq 10^{61}$~cm$^{-3}$ of
the latter is consistent with this, as the likely radius $R
\sim 10^{15}$~cm of this region (Begelman et al., 1980) implies $N_e
\sim 10^8$~cm$^{-3}$, and thus outflow rates as high as
\be \dot M_{\rm out}
\simeq 4\pi R^2vN_e m_H \sim 2.8\times 10^{-3}\ \msun{\rm yr}^{-1},
\label{10}
\ee
where $v \sim 1000$~km~s$^{-1}$ is the velocity width of the H$\alpha$
line.

The very high mass transfer rates encountered in these calculations
make it difficult to follow them to their natural endpoints, as
assumptions such as synchronous rotation of the donor begin to break
down. However the main outlines of the future evolution of systems
are fairly clear, provided that the system does not enter a CE phase
as a result of a delayed transition from thermal to dynamical
timescale mass transfer. (This delayed dynamical instability is avoided
in higher mass systems with mass ratio close to unity, again tending
to favor black--hole systems. Sequence 5 is likely to encounter this
instability, cf King \& Ritter, 1999). 
An initial mass ratio $q \ga 1$ implies that the Roche lobe
will shrink before the mass ratio reverses, and thus that
the current mass transfer phase is likely to end with a fairly tight
system consisting of the black hole or neutron star accretor
(mass effectively unchanged)
and the helium core of the donor (cf King \& Ritter, 1999).
For $M_2 < 12\msun$ at the onset of mass transfer the core has mass
$M_{\rm He} \sim 2\msun$. This star will
re--expand through helium shell--burning, and will probably initiate a
further
short mass transfer phase (so--called Case BB; Delgado \& Thomas, 1981;
Law \& Ritter, 1983), depending on the
binary separation. The donor will end its life as a CO white
dwarf. The binary may be tight
enough for coalescence to occur because of gravitational radiation
losses within $10^{10}$~yr. Systems of this type would therefore be
good candidates for gamma--ray burst sources, although detailed
evolutionary calculations to check the scenario sketched here are
clearly required before we can make this statement with any
confidence. In particular, it would be premature to translate the
predicted birthrates $10^{-4} - 10^{-5}$~yr$^{-1}$ into a predicted
gamma--ray burst rate.

\acknowledgements

This research was begun at the Institute for Theoretical Physics
and supported in part by the National Science Foundation under Grant
No. PHY94--07194. ARK gratefully acknowledges support by the UK
Particle Physics and Astronomy Research Council through a Senior
Fellowship. RT acknowledges support from NSF grant AST97--27875
MCB acknowledges support from NSF grants AST95--29170,
AST98--76887 and a Guggenheim Fellowship.

\begin{table*}
\begin{minipage}{\hsize}

\caption{\centerline{Possible Evolutionary Sequences for SS~433}
}
\bigskip

\begin{tabular}{ccccccc}
Sequence  & $t_0$ (yr)  & $M_1\ (\msun)$ & $M_{2i}\ (\msun)$ &
$M_{2}\ (\msun)$
& $\dot M_{\rm tr}\ (\msun\ {\rm yr}^{-1})$ & $P$ (d) \\

1 & 1007 & 6 & 8 & 7.990 & $2.3\times 10^{-5}$ & 13.09 \\

2 & 1003 & 4 & 8 & 7.987 & $3.3\times 10^{-5}$ & 13.05 \\

3 &  984 & 2 & 8 & 7.957 & $1.8\times 10^{-4}$ &  12.59 \\

4 & 1004 & 6 & 12 & 11.879 & $3.3\times 10^{-4}$ & 12.88 \\

5 & 1008 & 1.4 & 5 & 4.996 & $6.7\times 10^{-6}$ & 13.03 \\

\end{tabular}
\end{minipage}
\end{table*}

\end{document}